\documentclass[twocolumn,aps,showpacs]{revtex4}

\usepackage{graphicx}
\usepackage{epsfig}
\usepackage{dcolumn}
\usepackage{amsfonts}
\usepackage{amsmath}
\usepackage{amssymb}
\usepackage{bm}
\usepackage{longtable}

\begin{document}

%shortcuts
%\newcommand{\tens}[1]{\stackrel{\mbox{{\tiny $\leftrightarrow$}}}{#1}}
\newcommand{\tens}[1]{\tensor{#1}}
\newcommand{\brm}[1]{\bm{{\rm #1}}}

\title{Multifractality in directed percolation}

\author{Olaf Stenull}
\author{Hans-Karl Janssen}
\affiliation{
Institut f\"{u}r Theoretische Physik 
III\\Heinrich-Heine-Universit\"{a}t\\Universit\"{a}tsstra{\ss}e 1\\
40225 D\"{u}sseldorf\\
Germany
}

\date{\today}

\begin{abstract}
Using renormalization group methods we study multifractality in directed percolation. Our approach is based on random lattice networks consisting of resistor like and diode like bonds with microscopic noise. These random resistor diode networks capture the features of isotropic as well as directed percolation. In this note we introduce a field theoretic Hamiltonian for the multifractal properties at the transition from the non-percolating to the directed percolating phase. We investigate the multifractal moments of the current distribution and determine a family of critical exponents for these moments to two-loop order.
\end{abstract}

\maketitle

\noindent Nature is replete with systems where the exhaustive characterization of the distribution of a local physical quantity requires the introduction of an infinite set of independent critical exponents. For these systems the concept of multifractality~\cite{reviews_multifractality} has been developed. Examples for multifractal systems include turbulence~\cite{mandelbrot_74}, diffusion near fractals~\cite{cates_witten_87}, electrons in disordered media~\cite{wegner_87}, polymers in disordered media~\cite{gersappe_etal_91}, random ferromagnets~\cite{ludwig_87}, chaotic dissipative systems~\cite{halsey_etal_86}, random resistor networks~\cite{rammal_etal_85,arcangelis_etal_85,park_harris_lubensky_87,wir,stenull_janssen_2001,stenull_2000}, and heartbeat~\cite{ivanov_99}.

Here, we discuss multifractality in the context of directed percolation (DP)~\cite{hinrichsen_2000}. Consider two terminal points $x$ and $x^\prime$ on a fractal DP cluster. A current $I$ is inserted at $x$ and withdrawn at $x^\prime$. Particularly interesting quantities are the moments
\begin{eqnarray}
M_I^{(l)} \sim \sum_{\underline{b}} \left( I_{\underline{b}}/I \right)^{2l}
\end{eqnarray}
of the currents $I_{\underline{b}}$ in all the bonds $\underline{b}$ on the cluster. Specifically, $M_I^{(0)}$, $M_I^{(1)}$, $M_I^{(2)}$, and $M_I^{(\infty )}$ are proportional to the number of bonds on the backbone (bonds which carry nonzero current), the total resistance, the noise (second cumulant of the resistance fluctuations), and the number of the red bonds (bonds which carry the full current). If one measures $M_I^{(l)}$ in the direction along which the cluster is oriented, one finds that
\begin{eqnarray}
M_I^{(l)} \sim \big| x_\parallel - x_\parallel^\prime \big|^{\psi_l/\nu_\parallel} \ ,
\end{eqnarray}
where $| x_\parallel - x_\parallel^\prime |$ is the distance between the two terminals and $\nu_\parallel$ is the longitudinal correlation length exponent of the DP universality class. The multifractality manifests itself in the nonlinear dependence of $\psi_l$ on $l$. This behavior is much richer and more interesting than the usual gap scaling commonly encountered in critical phenomena. Of course, it is a challenging task to characterize the current distribution because one has to determine a multitude of critical exponents $\psi_l$. 

In this paper we study multifractality in DP based on the random resistor diode network (RDN) introduced by Redner~\cite{red_81_82_83}. Guided by ideas of Stephen~\cite{stephen_78}, Harris~\cite{harris_87}, and Park, Harris and Lubensky (PHL)~\cite{park_harris_lubensky_87} we introduce a field theoretic Hamiltonian for RDN with noise. We determine the multifractal exponents $\psi_l$ by computing the scaling indices of noise cummulants $C_R^{(l)}$ which are known to be proportional (cf.\ Ref.~\cite{blumenfeld&co_87}) to the $M_I^{(l)}$ by virtue of Cohn's theorem~\cite{cohn_50}. A central role in this analysis is played by our concept of master operators~\cite{wir,stenull_janssen_2001,stenull_2000}.

A RDN consists of a $d$-dimensional hypercubic lattice in which nearest-neighbor sites are connected by a resistor, a positive diode (conducting only in a distinguished direction), a negative diode (conducting only opposite to the distinguished direction), or an insulator with respective probabilities $p$, $p_{+}$, $p_{-}$, and $q=1-p-p_{+}-p_{-}$. In the three dimensional phase diagram (pictured as a tetrahedron spanned by the four probabilities) one finds a nonpercolating and three percolating phases~\cite{red_81_82_83}. The percolating phases are isotropic, positively directed, or negatively directed. Between the phases there are surfaces of continuous transitions. All four phases meet along a multicritical line, where $0\leq r:=p_{+}=p_{-}\leq 1/2$ and $p=p_{c}(r)$. On the entire multicritical line, i.e., independently of $r$, one finds the scaling properties of usual isotropic percolation ($r=0$). For the crossover from IP to DP see, e.g., Ref.~\cite{janssen_stenull_2000}. In this paper we focus on the vicinity of the critical surface separating the non-percolating and the positively directed phase. Here, typical clusters are anisotropic and they are characterized by two different correlation lengths: $\xi_{\parallel}$ (parallel to the distinguished direction) and $\xi_\perp$ (perpendicular to it). As one approaches the critical surface, the two correlation lengths diverge with the exponents $\nu_\parallel$ and $\nu_\perp$ of the DP universality class.

We choose ${\rm{\bf n}} = 1/\sqrt{d} \left( 1, \dots , 1 \right)$ for the distinguished direction, and we assume that the bonds $\underline{b}_{\langle i,j \rangle}$ between two nearest neighboring sites $i$ and $j$ are directed so that $\underline{b}_{\langle i,j \rangle} \cdot {\rm{\bf n}} > 0$. The directed bonds obey the non-linear Ohm's law $\sigma_{\underline{b}} \left( V_{\underline{b}} \right) V_{\underline{b}}  = I_{\underline{b}}$, where $V_{\underline{b}}$ is the potential drop over the bond and $I_{\underline{b}}$ denotes the current flowing from $j$ to $i$. The bond conductances $\sigma_{\underline{b}} = \varsigma_{\underline{b}} \, \gamma_{\underline{b}}$ are equally and independently distributed random variables. The $\gamma_{\underline{b}}$ take on the values $1$, $\theta \left( V \right)$, $\theta \left( -V \right)$, and $0$ with respective probabilities $p$, $p_+$, $p_-$, and $q$. $\theta$ denotes the Heaviside function. Static noise is introduced by distributing the $\varsigma_{\underline{b}}$ according to some distribution function $f$ with mean $\overline{\varsigma}$ and higher cumulants $\Delta^{(l\geq 2)}$ satisfying $\Delta^{(l)} \ll \overline{\varsigma}^l$. The noise average will be denoted by $\{ \cdots \}_f = \int \prod_{\underline{b}} d \varsigma_{\underline{b}} \, f \left( \varsigma_b \right) \cdots$ and its $l$th cumulant by $\{ \cdots^l \}^{(c)}_f$. Both kinds of disorder, the random dilution of the lattice and the noise, influence the statistical properties of the resistance $R_+ (x ,x^\prime )$~\cite{footnote1} between two sites $x$ and $x^\prime$. They are reflected by the noise cumulants 
\begin{eqnarray}
\label{defCumulant}
C_R^{(l)}(x ,x^\prime)
 = \frac{\big\langle \chi_+ (x ,x^\prime) \big\{ R_+ (x 
,x^\prime )^l \big\}_f^{(c)} \big\rangle_C}{\left\langle \chi_+ (x ,x^\prime) \right\rangle_C} \,  , 
\end{eqnarray}
where $\langle \cdots \rangle_C$ denotes the average over the configurations $C$ of the randomly occupied bonds and $\chi_+ (x ,x^\prime)$ is an indicator function which is unity if $x$ and $x^\prime$ are positively connected (i.e., if current can percolate from $x$ to $x^\prime$) and zero otherwise. 

Now we determine $\psi_l$ by exploiting the proportionality of $M_I^{(l)}$ and $C_R^{(l)}$. Our actual strategy is the following: (i) we devise a generating function for the $C_R^{(l)}$, (ii) we calculate this generating function by renormalized field theory, and (iii) we extract $C_R^{(l)}$ from the generating function.

To facilitate the averaging procedure we employ the replica technique~\cite{stephen_78}. Following PHL we replicate the voltages $D \hspace{-1mm}\times \hspace{-1mm}E$-fold, $V_x \to \tens{V}_x = ( V_x^{(\alpha ,\beta)})_{\alpha ,\beta = 1}^{D,E}$. In order to set up the desired generating function we introduce $\psi_{\tens{\lambda}}(x) = \exp ( i \tens{\lambda} \cdot \tens{V}_x )$, where $\tens{\lambda} \cdot \tens{V}_x = \sum_{\alpha , \beta =1}^{D,E} \lambda^{(\alpha , \beta )} V_x^{(\alpha , \beta)}$ and $\tens{\lambda} \neq \tens{0}$ and define corresponding correlation functions 
\begin{eqnarray}
\label{noisyErzeugendeFunktion}
\lefteqn{ G \big( x, x^\prime ,\tens{\lambda} \big)
=\Bigg\langle \Bigg\{ \frac{\int \prod_j \prod_{\alpha ,\beta=1}^{D,E} 
dV_j^{(\alpha , \beta )} }{\prod_{\beta =1}^E Z ( \{ \sigma_b^{(\beta )} \} , C )^D } 
}
\nonumber \\
& & \times \exp \bigg[ -\frac{1}{2} P \big( \big\{  \vec{V} \big\} \big) + i \tens{\lambda} \cdot \big( \tens{V}_x  - \tens{V}_{x^\prime} 
\big) \bigg] \Bigg\}_f \Bigg\rangle_C  .
\end{eqnarray}
Here $P ( \{ \tens{V} \} ) = \sum_{\alpha ,\beta =1}^{D,E} \sum_{\underline{b}} \varsigma_{\underline{b}}^{(\beta )} \, \gamma_{\underline{b}} ( V_{\underline{b}}^{(\alpha ,\beta)} )  V_{\underline{b}}^{(\alpha ,\beta) 2}$ is the replicated power dissipated on the backbone and $Z$ is the usual normalization.

Due to the appearance of the $\theta$-functions in the power $P$ the integration in Eq.~(\ref{noisyErzeugendeFunktion}) is not Gaussian. However, it can be carried out in an approximating manner by applying the saddle point method as it was done by Harris~\cite{harris_87} in the related problem of nonlinear random resistor networks. We extract the leading contribution to the integral steaming from the maximum of the integrand. This maximum is determined by the solution of Kirchhoff's equations, i.e., by the resistance $R_+ (x,x^\prime)$. Hence one finds (details on the steps can be gleaned from Refs.~\cite{harris_87,park_harris_lubensky_87}) that
\begin{eqnarray}
\label{cumulantGenFkt}
\lefteqn{ G \big( x, x^\prime ,\tens{\lambda} \big) \sim }
\nonumber \\
&&
\bigg\langle  \exp \bigg[ \sum_{l=1}^\infty \frac{(-1/2)^l}{l!} K_l \big( 
\tens{\lambda} \big) \big\{ R_+ 
\left( x,x^\prime \right)^l \big\}_f^{(c)} \bigg]  \bigg\rangle_C , 
\end{eqnarray}
where $K_l ( \tens{\lambda} ) = \sum_{\beta =1}^E [ \sum_{\alpha =1}^D ( \lambda^{(\alpha ,\beta )} )^2 ]^l$. Equation~(\ref{cumulantGenFkt}) shows that $G$ represents the desired generating function from which $C_R^{(l)}$can be calculated via taking the derivative with respect to $K_l (\tens{\lambda})$.

A glance at Eq.~(\ref{noisyErzeugendeFunktion}) reveals the benefit of the replication. It provides an additional parameter $D$ which we can tune to zero. In this replica limit the normalization denominator in Eq.~(\ref{noisyErzeugendeFunktion}) is taken to unity and in particular it no longer depends on the disorder. Hence we are left with the much simpler task to average the exponential in Eq.~(\ref{noisyErzeugendeFunktion}) which gives us an effective Hamiltonian 
\begin{eqnarray}
H_{\mbox{\scriptsize{rep}}} =  - \ln \bigg\langle \left\{ \exp \left[ - \frac{1}{2} P \big( \big\{ \tens{V} \big\} \big) \right] \right\}_f \bigg\rangle_C .
\end{eqnarray}

For technical reasons~\cite{park_harris_lubensky_87,wir,stenull_janssen_2001,stenull_2000} we switch to discretized voltages $\tens{\vartheta}$ and currents $\tens{\lambda}$ taking values on a discrete $D \times E$-dimensional torus. For the saddle point method to be reliable we work near the limit when all the components of $\tens{\lambda}$ are equal and continue to large imaginary values. Accordingly we set~\cite{harris_87} $\lambda^{(\alpha ,\beta )} = i \lambda_0 + \xi^{(\alpha ,\beta )}$ with real $\lambda_0$ and $\xi^{(\alpha ,\beta )}$, $\sum_{\alpha ,\beta =1}^{D,E} \xi^{(\alpha ,\beta )} = 0$, and impose the conditions $\lambda_0^{2} \ll D^{-1}$ and $\tens{\xi}^2 \ll 1$.

To refine $H_{\mbox{\scriptsize{rep}}}$ towards a field theoretic Hamiltonian, we expand $H_{\mbox{\scriptsize{rep}}}$ in terms of $\psi_{\tens{\lambda}} \left( x \right)$. The steps are analogous to those in Ref.~\cite{harris_87} and are skipped here for briefness. The so obtained expression is converted into a Landau-Ginzburg-Wilson-type functional
\begin{eqnarray}
\label{hamiltonian}
{\mathcal{H}} \hspace{-1.5mm} &=& \hspace{-1.5mm} \int d^dx \bigg\{ \frac{1}{2} \sum_{\tens{\lambda} \neq \tens{0}} \psi_{-\tens{\lambda}} \left( {\rm{\bf x}} \right) \bigg[  \tau - \nabla^2 + w \tens{\lambda}^2 
\nonumber \\
&+& \hspace{-1.5mm} \sum_{l=2}^{\infty} v_l K_l \left( \tens{\lambda} \right)
+ \left[ \theta \left( \lambda_0 \right) - \theta \left( -\lambda_0 \right) \right] {\rm{\bf v}} \cdot \nabla \bigg] \psi_{\tens{\lambda}} \left( {\rm{\bf x}} \right)
\nonumber \\
&+& \hspace{-1.5mm} \frac{g}{6} \sum_{\tens{\lambda}, \tens{\lambda}^\prime  , \tens{\lambda} + \tens{\lambda}^\prime \neq \tens{0}} \psi_{-\tens{\lambda}} \left( {\rm{\bf x}} \right) \psi_{-\tens{\lambda}^\prime} \left( {\rm{\bf x}} \right) \psi_{\tens{\lambda} + \tens{\lambda}^\prime} \left( {\rm{\bf x}} \right) \bigg\}
\end{eqnarray}
by applying the usual coarse graining procedure. The parameter $\tau$ specifies the ``distance'' from the critical surface under consideration. The vector ${\rm{\bf v}}$ lies in the distinguished direction, ${\rm{\bf v}} = {\rm v} {\rm{\bf n}}$. $\tau$ and $v$ depend on the three probabilities $p$, $p_+$, and $p_-$. $w$ and $v_l$ are the coarse grained analogs of $\overline{\varsigma}^{-1}$ and $\Delta^{(l)} / \overline{\varsigma}^{2l}$, respectively. In the limit $w\to 0$ and $v_l \to 0$ our Hamiltonian ${\mathcal{H}}$ describes the usual purely geometric DP. Indeed ${\mathcal{H}}$ leads for $w\to 0$ and $v_l \to 0$ to exactly the same perturbation series as obtained in Refs.~\cite{cardy_sugar_80,janssen_81,janssen_2000}. 

Now we address the relevance of the $v_l$. A straightforward scaling analysis reveals that
\begin{eqnarray}
\label{cumulantScaling}
\lefteqn{
C_R^{(l)} \left( \left( {\rm{\bf x}}, {\rm{\bf x}}^\prime \right) ; \tau , {\rm v}, w, \left\{ v_l \right\} \right)
}
\nonumber \\
&& = w^l f_l \left( \left( {\rm{\bf x}}, {\rm{\bf x}}^\prime \right) ; \tau , {\rm v}, \left\{ \frac{v_k}{w^k} \right\} \right)  ,
\end{eqnarray}
where $f_l$ is a scaling function. Note that the coupling constants $v_k$ appear only as $v_k / w^k$. Dimensional analysis shows that $w \tens{\lambda}^2 \sim \mu^2$ and  $v_k K_k \big( \tens{\lambda} \big) \sim \mu^2$, where $\mu$ is the usual inverse length scale. Thus, $v_k / w^k \sim \mu^{2-2k}$, i.e., the $v_k / w^k$ have a negative naive dimension which decreases drastically with increasing $k$. We conclude that the $v_{k}$ are irrelevant couplings. Though irrelevant, we must not set $v_{k}=0$ in calculating the multifractal exponents. To see this we expand the scaling function $f_l$ yielding
\begin{eqnarray}
\label{expOfCumulantScaling1}
\lefteqn{
C_R^{(l)} \left( \left( {\rm{\bf x}}, {\rm{\bf x}}^\prime \right) ; \tau , {\rm v}, w, \left\{ v_l \right\} \right)
}
\nonumber \\
&& = v_l \Big\{ C_l^{(l)} + C_{l+1}^{(l)} \frac{v_{l+1}}{w \, v_l} + \cdots \Big\} ,
\end{eqnarray}
with $C_k^{(l)}$ being an expansion coefficient depending on ${\rm{\bf x}}$, ${\rm{\bf x}}^\prime$, $\tau$, and ${\rm v}$. Note that $C_{k<l}^{(l)} = 0$ because the corresponding terms are not generated in the perturbation calculation. The first term on the right hand side of Eq.~(\ref{expOfCumulantScaling1}) gives the leading behavior. Thus, $C_R^{(l)}$ vanishes upon setting $v_l = 0$ and we cannot gain any further information about $C_R^{(l)}$. In other words, the $v_{l}$ are dangerously irrelevant in investigating the critical properties of the $C_R^{(l\geq 2)}$.

We apply standard methods of field theory~\cite{amit_zinn-justin} and perform a diagrammatic perturbation calculation up to two-loop order. Since the coupling constants $v_l$ are irrelevant, they cannot be treated in the same fashion as the other coupling constants $\tau$ and $w$ pertaining to the bilinear part of $\mathcal{H}$. Using $[\tau + {\rm{\bf p}}^2 + w \vec{\lambda}^2 + \sum_l v_l K_l ( \tens{\lambda})]^{-1}$ as the Gaussian propagator would poison the Feynman diagrams. This can be understood by expanding any given diagram in terms of $v_l$. For increasing orders of this expansion one encounters increasing orders of primitive divergence. Hence one has to truncate this expansion at linear order in $v_l$ which is equivalent to treating $v_l$ by means of the insertion 
\begin{eqnarray}
\label{opdev}
\mathcal{O}^{(l)} = - \frac{1}{2} v_l \int d^d p \, \, {\textstyle \sum_{\tens{\lambda}} } K_l \big( \tens{\lambda} \big) \phi_{\tens{\lambda}} \left( {\rm{\bf p}} \right) \phi_{-\tens{\lambda}} \left( -{\rm{\bf p}} \right)  ,
\end{eqnarray}
where $\phi_{\tens{\lambda}} \big( {\rm{\bf p}} \big)$ denotes the Fourier transform of $\psi_{\tens{\lambda}} \big( {\rm{\bf x}} \big)$. This insertion does not only generate primitive divergencies proportional to $K_l ( \tens{\lambda} )$, but also primitive divergences corresponding to all operators $\mathcal{O}^{(l)}_i$ of the generic form $\tens{\lambda}^{2a} {\rm{\bf p}}^{2b} \phi^n$ having the same or a lower naive dimension than $\mathcal{O}^{(l)}$. In other words: all operators of the form $\tens{\lambda}^{2a}{\rm{\bf p}}^{2b} \phi^n$ with $a+b+n \leq l+2$ are generated. Those $\mathcal{O}^{(l)}_i$ associated with a lower naive dimension than $\mathcal{O}^{(l)}$ can be neglected because they just lead to corrections which vanish at the critical point (cf.\ Ref.~\cite{amit_zinn-justin}). Anyway, one still needs a myriad of operators as counterterms in the Hamiltonian. Inserting either of these $\mathcal{O}^{(l)}_i$, however, does not generate $\mathcal{O}^{(l)}$ because the $\mathcal{O}^{(l)}_i$ possess of higher symmetries than $\mathcal{O}^{(l)}$. Thus, the renormalization scheme is given by
\begin{eqnarray}
\mathcal{O}^{(l)}_{\mbox{\scriptsize bare \normalsize}} = Z^{(l)} \, \mathcal{O}^{(l)}_{\mbox{\scriptsize ren \normalsize}} + \sum_i Z^{(l)}_i \, \mathcal{O}^{(l)}_{i, \hspace{1mm}\mbox{\scriptsize ren \normalsize}} ,
\\
\mathcal{O}^{(l)}_{i, \hspace{1mm}\mbox{\scriptsize bare \normalsize}} = \sum_i Z^{(l)}_{i,j} \, \mathcal{O}^{(l)}_{j, \hspace{1mm}\mbox{\scriptsize ren 
\normalsize}} ,
\end{eqnarray}
and one solely needs $Z^{(l)}$ in calculating the scaling index of $\mathcal{O}^{(l)}_{\mbox{\scriptsize ren \normalsize}}$. What we encounter here is another realization of our concept of master operators~\cite{wir,stenull_janssen_2001,stenull_2000}. Each multifractal moment $M_I^{(l)}$ corresponds one-to-one to a dangerously irrelevant master operator $\mathcal{O}^{(l)}$. Though a myriad of servant operators is involved in the renormalization of the masters $\mathcal{O}^{(l)}$ the scaling behavior of $M_I^{(l)}$ is governed by $\mathcal{O}^{(l)}$ only. The servant operators can be neglected in determining the scaling index of their master operator, i.e., we are spared the computation and diagonalization of giant renormalization matrices.

By employing dimensional regularization and minimal subtraction we proceed with standard techniques of renormalized field theory. We calculate $Z^{(l)}$ to two-loop order. From the renormalization group equation and dimensional analysis we deduce that the correlation function $G$ scales at criticality as
\begin{eqnarray}
\label{corrfkt}
\lefteqn{
G \left( {\bf 0}, {\bf x} ,\tens{\lambda} \right) 
= 
x_\parallel^{(1-d-\eta )/z} f \left(  \frac{\left| {\rm{\bf x}}_\perp \right|^z}{x_\parallel} \right) 
 \bigg\{ 1 
+ w \vec{\lambda}^2 x_\parallel^{\phi /\nu_\parallel} 
}
\nonumber \\
&\times& \hspace{-1.5mm}
f_w \left(  \frac{\left| {\rm{\bf x}}_\perp \right|^z}{x_\parallel} \right)+ 
 v_l K_l \left( \tens{\lambda} \right) x_\parallel^{\psi_l /\nu_\parallel} f_v \left(  \frac{\left| {\rm{\bf x}}_\perp \right|^z}{x_\parallel} \right) \bigg\} ,
\end{eqnarray}
where higher order terms have been discarded. $\eta$, $\nu_\parallel$, and $z= \nu_\parallel /\nu_\perp$ are the critical exponents for DP known to second order in $\epsilon =5-d$~\cite{janssen_81,janssen_2000}. $f$, $f_w$, and $f_v$ are scaling functions. $\phi$ is the resistance exponent given to second order in $\epsilon$ in Ref.~\cite{janssen_stenull_directedLetter_2000}. In Eq.~(\ref{corrfkt}) we introduced the multifractal exponents $\psi_l = \nu_\perp \left( 2 - \gamma^{(l)\ast} \right)$, where $\gamma^{(l)\ast}$ stands for the Wilson function $\gamma^{(l)} = - \mu \frac{\partial}{\partial \mu} \ln Z^{(l)} |_0$ evaluated at the infrared stable fixed point. By taking the derivative with respect to $K_l (\tens{\lambda})$ we can now deduce the scaling behavior of the multifractal moments: $M_I^{(l)} \sim x_\parallel^{\psi_l /\nu_\parallel}$ if measured parallel to the distinguished direction.  For measurements in other directions it is appropriate to choose a length scale $L$ and to express the longitudinal and the transverse coordinates in terms of $L$: $\left| {\rm{\bf x}}_\perp \right| \sim L$ and $x_\parallel \sim L^z$. With this choice we find that $M_I^{(l)} \sim L^{\psi_l /\nu_\perp}$. 

The result for the multifractal exponents remains to be stated. For $l\geq 0$ we obtain to second order in $\epsilon$
\begin{eqnarray}
\label{monsterExponent}
\psi_l = 1 + \frac{\epsilon}{3 \cdot 2^{2l+1}} + \epsilon^2 \left[ a(l) - b(l) \, \ln \left( \frac{4}{3} \right) \right] ,
\end{eqnarray}
with $a(l)$ and $b(l)$ taking on the values listed in Table~\ref{tab:coeffs}.
%%%%table%%%%%
%\begin{widetext}
\begin{table*}[p]
\caption{The coefficients $a(l)$ and $b(l)$ appearing in Eq.~\ref{monsterExponent}.}
\label{tab:coeffs}
%\begin{ruledtabular}
\begin{tabular}{c||c|c|c|c|c|c|c|c}
\hline \hline
$\quad l \quad $ & $0$ & $1$ & $2$ & $3$ & $4$ & $5$ & $6$ & $\geq 7$\\ \hline
$ a(l) $ & $\frac{85}{1728}$ & $\frac{151}{6912}$ & $\frac{68387}{4976640}$ & $\frac{3307921}{334430208}$ & $\frac{4661703289}{619173642240}$ & $\frac{8258257317517}{1373079469031424} $ & $\frac{24071498466367}{4808089723207680}$ & $0.005 > a > 0$\\
\hline
$ b(l) $ & $-\frac{53}{864}$ & $\frac{157}{3456}$ & $\frac{1091}{27648}$ & $\frac{13589}{442368}$ & $\frac{173149}{7077888}$ & $\frac{2281853}{113246208}$ & $\frac{30950909}{1811939328}$ & $0.015 > b > 0$\\
\hline \hline
\end{tabular}
%\end{ruledtabular}
\end{table*}
%\end{widetext}
%%%%%%%%%%%%%
$\psi_0$ and $\psi_1$ stem from extending the sum over $l$ in Eq.~(\ref{hamiltonian}) to comprise $l=0$ and $l=1$. Equation~(\ref{monsterExponent}) fulfills several consistency checks. $\psi_0$ is related to the fractal dimension $D_B$ of DP clusters via $D_B = 1 + \psi_0 /\nu_\perp$. Equation~(\ref{monsterExponent}) is in agreement with the $\epsilon$-expansions of $\nu_\perp$~\cite{janssen_81,janssen_2000} and $D_B$ (cf.\ Ref.~\cite{janssen_stenull_directedLetter_2000}) to second order in $\epsilon$. $\psi_1$ is in conformity with our result for the resistance exponent $\phi$ given in Ref.~\cite{janssen_stenull_directedLetter_2000,stenull_janssen_2001_jsp}. As anticipated $\psi_l$ is for reasonable values of $\epsilon$ a convex monotonically decreasing function of $l$~\cite{duplantier_ludwig_91}. It tends to unity for large $l$ as one expects from the relation of $\psi_\infty$ to the fractal dimension of the singly connected (red) bonds, cf.\ Refs.~\cite{arcangelis_etal_85,coniglio_81_82}.

While directed percolation is a vivid area of research since about 20 years and multifractality in ordinary percolation is studied since about 15 years, the problem of multifractality in DP has not been addressed hitherto, at least to our knowledge. In the present note, we introduced a field theoretic Hamiltonian for the multifractal properties of resistor diode percolation. We applied our concept of master operators. It turned out that it works consistently as a tool to describe the multifractal properties of DP by renormalized field theory. Without the concept of master operators which avoids the calculation and diagonalization of giant renormalization matrices the actual calculations are hardly feasible. We introduced the family of multifractal exponents $\{ \psi_l \}$ in the spirit of PHL. We calculated $\psi_l$ for non-negative $l$ to two-loop order. Our result fulfills several consistency checks. It is certainly desirable to have numerical data for comparison with our analytic results, in particular in dimensions to which the $\epsilon$-expansion can be continued reliably. We hope that this paper triggers such simulations.

We acknowledge support by the Sonderforschungsbereich 237 ``Unordnung und gro{\ss}e Fluktuationen'' of the Deutsche Forschungsgemeinschaft. 

%references


\begin{thebibliography}{}
\bibitem{footnote1}
The subscript signals that we are interested in the forward resistance. Equally well one could study the backward resistance $R_- (x ,x^\prime ) = R_+ (x^\prime ,x )$.
\end{thebibliography}
\end{document}